\documentstyle[epsf,epsfig,preprint,aps,amssymb]{revtex}
\draft \preprint{KIAS-P02003; SNUTP 02/007}
\begin{document}
\title{\Large \bf TeV scale 5D $SU(3)_W$ unification and the
fixed point anomaly cancellation with chiral split multiplets }
\author{$^{(a)}$Hyung Do Kim\footnote{\tt hdkim@kias.re.kr},
$^{(b)}$Jihn E. Kim\footnote{\tt jekim@phyp.snu.ac.kr} and
$^{(b)}$Hyun Min Lee\footnote{\tt minlee@phya.snu.ac.kr}}
\address{$^{(a)}$School of Physics, Korea Institute for Advanced
Study, Cheongryangri-dong, Dongdaemun-ku, Seoul 135-012, Korea\\
$^{(b)}$School of Physics and Center for Theoretical Physics,
Seoul National University, Seoul 151-747, Korea}
\maketitle

\def\bea{\begin{eqnarray}}
\def\eea{\end{eqnarray}}
\def\nn{\nonumber}
\def\g{g^{'}}
\def\bi{\tilde{b_i}}
\setlength{\textwidth}{17.0cm} \setlength{\textheight}{22.0cm}
\setlength{\oddsidemargin}{-0.1cm}
\setlength{\evensidemargin}{0.5cm} \setlength{\headheight}{0cm}
\setlength{\headsep}{0cm} \setlength{\topmargin}{0cm}
\setlength{\footskip}{1.0cm}

\baselineskip 0.6cm \tighten \vskip 0.5cm

\abstract{A possibility of 5D gauge unification of $SU(2)_L \times
U(1)_Y$ in $SU(3)_W$ is examined. The orbifold compactification
allows fixed points where $SU(2)_L\times U(1)_Y$ representations
can be assigned. We present a few possibilities which give long
proton lifetime, top-bottom mass hierarchy from geometry, and
reasonable neutrino masses. In general, these {\it chiral models}
can lead to fixed point anomalies. We can show easily, due to the
simplicity of the model, that these anomalies are cancelled by the
relevant Chern-Simons terms for all the models we consider. It is
also shown that the fixed point $U(1)$--graviton--graviton anomaly
cancels without the help from the Chern-Simons term. Hence, we
conjecture that the fixed point anomalies can be cancelled if the
effective 4D theory is made anomaly free by locating chiral
fermions at the fixed points.\\ \noindent [Key words: TeV scale
unification, proton stability, split multiplets, orbifolds,
anomaly cancellation]}


\newpage

\section{Introduction}

It is believed that the supersymmetric extension of the standard
model predicts a unification of gauge coupling constants with the
desert hypothesis. In the minimal supersymmetric standard model
(MSSM) three gauge couplings meet within the experimental error
bound at $M_{\rm GUT} = 2 \times 10^{16}$ GeV with a TeV
supersymmetry breaking scale \cite{dg,drw}. This scenario involves
large logarithms necessary to make the observed $\alpha_3$ large
and $\sin^2 \theta_W\simeq 0.231$ at $M_Z$ starting from the
universal coupling at high energy, around $\alpha_{3,2,1}\simeq
1/25$ and $\sin^2\theta_W^0=3/8$. The wide desert is unavoidable
in this scenario and introduces the gauge hierarchy
problem\cite{gildener}.

There are other proposals for the solution of the gauge hierarchy
problem. The recent effort with the low energy(TeV scale)
fundamental scale with extra dimensions tries to solve it simply
by abolishing the large hierarchy\cite{Arkani-Hamed}. If, the
fundamental scale is indeed TeV scale, the TeV scale quantum
gravity can occur, which is very interesting by itself. Then the
conventional supersymmetric grand unified theory (SUSY GUT) should
be modified in this framework. Running of gauge couplings in the
extra dimensional model is different from the four dimensional
one\cite{Dienes} and reflects several interesting features. With
the extra dimension(s), the compactification down to 4 dimensional
space-time(4D) is a necessity. Starting with odd space-time
dimensions, there does not arise the gauge anomaly problem simply
because the spinor representations are real. However, the
compactification procedure produce complex fermions. Without some
twist of the internal space as in the simple torus
compactification, the Kaluza-Klein(KK) levels are left-right
symmetric, i.e. the fermion representation is vectorlike and there
are no massless fermions in the low energy 4D theory. Thus, to
have chiral fermions, we should twist the internal space
\cite{dixon,inq}. If the torus is modded out by a discrete group
$Z_N$, then there appear some fixed points. The level matching in
the bulk is shifted and the fermions in the bulk need not be
vectorlike as in the models studied in Refs.\cite{acg,bhn,kkl}.
Then, there should be chiral fermions at the fixed points so that
there is no anomaly in the effective 4D theory. However, there can
be fixed point anomalies\cite{acg}. We try to investigate the
possibility of TeV scale unification of gauge coupling constants
along this line.

Note that the $SU(5)$ unification starts from a large
$\sin^2\theta_W^0$ at the unification scale, which needed a wide
desert. For a TeV scale unificationin, i.e. in models without
desert, therefore, the bare $\sin^2\theta_W^0=1/[1+C^{2}]$ must be
close to 0.231, where $C^2$ defines a properly normalized $U(1)_Y$
hypercharge. In $SU(5)$, $C^2=5/3$ and
$\sin^2\theta_W^0=\frac{3}{8}$. If $C^2=3$, then $\sin^2\theta_W^0
= \frac{1}{4}$ which can allow TeV scale unification. This
$U(1)_Y$ normalization occurs if $SU(2)_L \times U(1)_Y$ is
unified in $SU(3)_W$ with the simplest embedding of ${\bf
2}_{\frac{1}{2}}$ of $SU(2)_L \times U(1)$ into ${\bf 3}$ of
$SU(3)_W$\cite{weinberg}. Though leptons and Higgs can be nicely
embedded into $SU(3)_W$, we encounter a difficulty of explaining
$\lq$fractional' hypercharges of quarks (in the unit of
$\frac{1}{2}$). Suppose $L=(\nu,e)$ and $e^c$ belong to ${\bf 3}$
of $SU(3)_W$. The hypercharges of $L$ and $e^c$ are normalized to
$-\frac{1}{2}$ and $1$, respectively and satisfies $Tr\ Y=0$ for
the ${\bf 3}$ of $SU(3)_W$. If $Q=(u,d)$ and $d^c$ come from ${\bf
3}$ of $SU(3)_W$ , then $Y_Q = \frac{1}{6}$ and
$Y_{d^c}=\frac{1}{3}$ need an extra $U(1)$ group beyond $SU(3)_W$,
in conflict with our motivation for predicting the weak mixing
angle from a unified theory.

Recently five dimensional(5D) models for gauge coupling
unification showed that incomplete multiplets can be consistently
introduced at the fixed point (SM fixed point) where only part of
the gauge symmetry survives after the orbifold breaking of the
gauge symmetry. Though nonuniversal gauge kinetic terms are
introduced in this case, the uncertainty coming from them at the
fixed point is suppressed by the large volume factor compared to
the universal kinetic term in the bulk. Therefore, we obtain a
reasonable unification relation for the low energy gauge coupling
constants as long as the volume of the extra dimension is large
enough. From the minimally deconstructed point of view, this model
can be interpreted as gauge group $G_{\rm GUT} \times G_{\rm SM\
High}$ where $G_{\rm GUT}$ is the bulk gauge group and $G_{\rm SM\
High}$ is the gauge group at the concerned fixed point. Largeness
of extra dimension is interpreted as the strong coupling of
$G_{\rm SM}$ since low energy gauge coupling after the breaking of
the gauge group to its diagonal $G_{\rm SM}={\rm
diag.}(G_{GUT}\oplus G_{\rm SM\ High})$ is given by
 \bea
 \frac{1}{g_i^2} & = & \frac{1}{g_{\rm GUT}^2} + \frac{1}{g_{{\rm
 SM\ High},i}^2}
 \eea
for the factor group $G_i$ of $G_{\rm SM}$. Thus in this setup,
the difficulty of obtaining correct $Y_Q$ can be easily avoided
once we assume that the quark multiplets live at the SM fixed
point. Recently there appeared series of papers allowing TeV scale
unification of the electroweak sector into $SU(3)_W$
\cite{DK,LW,HN,DKW,DK2}.

Now it would be meaningful to search for the possibility of
unification in various setups even though there are uncertainties
due to the absence of exact spectrum of Higgs and/or
superpartners.

It was shown that two gauge coupings meet if the inverse of the
extra dimension size is 1 $-$ 2 TeV for the SM matters (with one
Higgs doublet) . For supersymmetric case, the size of the extra
dimension is given by 3 TeV to 6 TeV \cite{HN}, and strong
coupling scale in which two couplings meet is two orders higher
and is about a few hundred TeV. The supersymmetric model in
\cite{HN} has two Higgses in the bulk and matter multiplets on the
fixed point. SUSY breaking scale is assumed to vary from $M_Z$ to
1 TeV and the result is quite sensitive to the spectrum of
superpartners. However, putting the matter fields in the bulk or
at the fixed points is in a sense arbitrary.

The arbitrariness of putting the matter fields is a bit regulated
if we adopt a $\lq$naturalness' scheme: If a GUT group is broken
at a fixed point, then put a multiplet at that fixed point allowed
by the representation of the fixed point gauge group \cite{HM}.
This was known in string orbifold compactifications\cite{iknq}.
Mainly the fixed point gauge group determines the configuration of
the model. One such example was given in Ref. \cite{kkl} in which
the $SU(5)$ gauge fields propagates in the bulk. The minimal setup
for Higgs is to put $H_d$ in the bulk. The orbifold
compactification leads to a chiral fermion in the bulk. So with
this representation the effective 4D theory is anomalous. One can
introduce $H_u$ also in the bulk. Then the theory is vectorlike
and there is no anomaly problem\cite{kawamura}. We call this {\it
vectorlike models}. On the other hand we can put $H_u$ at the
fixed point $A$. The fixed point $A$ has the gauge symmetry
$G_{\rm SM}\equiv SU(3)\times SU(2)\times U(1)$ and $G_{\rm SM}$
representations are allowed at the asymmetric fixed point $A$:
some chiral fermions and $H_u$. This kind of asymmetric embedding
of chiral fermions are called {\it chiral models.}

Similarly with the spirit of Ref.\cite{kkl}, we study the
$SU(3)_W$ models along this line in this paper. In Sec. II, we
present possible $SU(3)_W$ models and calculate the running of the
gauge coupling constants. In Sec. III, the proton decay problem is
addressed and solved by putting leptons at the {\it symmetric}
fixed point $O$ and quarks at the {\it asymmetric} fixed point
$A$. In Sec. IV, the neutrino masses are discussed. In Sec. V, we
show that fermionic anomalies at the fixed points are cured by the
Chern-Simons term in the bulk in these asymmetric models. Sec. VI
is a conclusion.

\section{Models}

The natural assignment of the representations at the fixed points
allows only the representations of the fixed point gauge group,
but not necessarily the full representation of the unbroken gauge
group. We define the {\it minimal model} in which the fermion
representation is put at the fixed point where the gauge group is
maximum. Let us consider 5D $SU(3)_W$ models.

In the bulk, the gauge fields propagate. In the $S_1/Z_2\times
Z_2^\prime$ compactification with the Sherk-Schwarz mechanism,
there appear two fixed points
\begin{eqnarray}
&O_{\rm GUT}& : \ y=0\
\nonumber\\
&A& : \ y=\frac{\pi}{2}R\  \nonumber
\end{eqnarray}
where $O$ is the {\it symmetric fixed point} where $SU(3)_W$ is
not broken, and $A$ is the {\it asymmetric fixed point} where
$SU(3)$ is broken down to $SU(2)_L\times U(1)_Y$. Let us put
fermions at the fixed points in the minimal scheme. The leptons
$L$ and $e^c$ can form an $SU(3)_W$ triplet {\bf 3}, and hence we
put it at the symmetric fixed point $O$. On the other hand, $Q$,
$u^c$ and $d^c$ cannot form $SU(3)_W$ representations. So they
must be put at the asymmetric fixed point $A$. The Higgs doublet
$H_d(Y=-\frac{1}{2})$ or $H_u(Y= \frac{1}{2})$ can be assigned
into ${\bf 3}_H$ or ${\bf \bar{3}}_H$ of $SU(3)_W$. Hence, if the
vacuum expectation value of  ${\bf 3}_H$ gives mass to electron,
it must be put where $SU(3)_W$ is a good symmetry, i.e. in the
bulk or at $O$. If this ${\bf 3}_H$ also gives mass to quark(s),
then it must be put in the bulk. Since it is more economic to have
less Higgs fields, the minimal model dictates to put ${\bf 3}_H$
in the bulk. On the other hand, we put the Higgs doublet $H_u$ at
the asymmetric fixed point $A$.

This minimal setup has several merits. First, rapid proton decay
can be avoided if the size of the extra dimension is slightly
bigger than the fundamental length since quarks and leptons are
located at the opposite fixed points and the locality of the extra
dimension prevents the contact term giving rapid proton decay.
Second, it gives a geometric explanation of $b-t$ mass
ratio\cite{kkl}.

With supersymmetry, there is another possibility for the fixed
point gauge symmetry {\it a la} the Sherk-Schwarz mechanism as
discussed in Sec. IV of \cite{HN}. When we compactify 5D on an
$S^1/(Z_2\times Z'_2)$, $Z_2$ and $Z'_2$ turn out to be equivalent
to one orbifolding $Z$ for $y\rightarrow -y$ and one twist $T$ for
$y\rightarrow y+\pi R$. So we conveniently write the boundary
conditions of the gauge field $A_M=(A_\mu,A_5)$ $(\mu=0,1,2,3)$ in
terms of $Z$ and $T$ :
\begin{eqnarray}
A_\mu(y)&=&Z A_\mu(-y) Z^{-1}=T A_\mu(y+\pi R) T^{-1} \\
A_5(y)&=&-Z A_5(-y) Z^{-1}=T A_5(y+\pi R) T^{-1}
\end{eqnarray}
where $Z$, $T$ and $A_M=A^a_M T^a$ are represented by $3\times 3$
matrices. Then there are two independent choices for ${Z,T}$ for
breaking $SU(3)_W$ into $SU(2)_L\times U(1)_Y$,
\begin{eqnarray}
{\rm Type\ I}:\ (Z,T)&=&({\rm diag}(1,1,1), {\rm
diag}(1,1,-1)),\nonumber
\\
{\rm Type\ II}: (Z,T)&=&({\rm diag}(1,1,-1), {\rm
diag}(1,1,-1)).\nonumber
\end{eqnarray}
For the former case, the $SU(3)_W$ symmetry breaks into $SU(2)_L
\times U(1)_Y$ only at the fixed point $A$ while it is fully
conserved at the fixed point $O$. On the other hand, for the
latter case, the $SU(3)_W$ symmetry is broken into $SU(2)_L\times
U(1)_Y$ at both fixed points. However, for Type II, there also
appears a massless adjoint scalar of $SU(2)_L\times U(1)_Y$, which
nonetheless gets a radiative mass of the order of the
compactification scale.

\vskip 0.3cm
\begin{figure}[b]
\centering \centerline{\epsfig{file=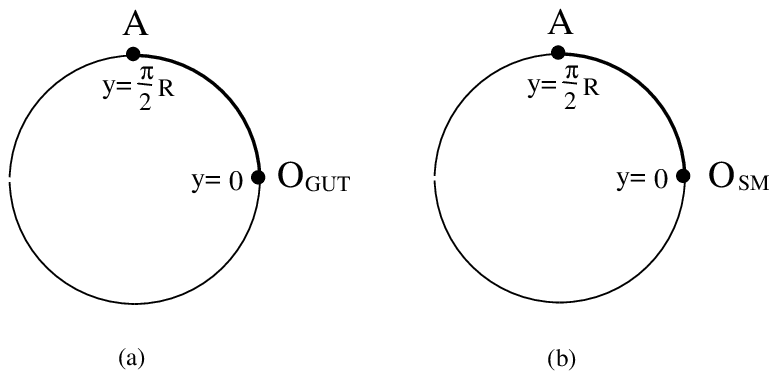,width=120mm}}
\end{figure}
\centerline{ Fig.~1.\ \it  The asymmetric point $A$ preserves only
the SM gauge group. On the other hand,} \centerline{\it the
symmetric point $O$ can ({\rm a}) preserve $SU(3)_W$, or preserve
only} \centerline{({b}){\it\  the SM gauge group, which are
explicitly shown as $O_{\rm GUT}$}} \centerline{\it and $O_{\rm
SM}$, respectively. In this paper, we consider only $O_{\rm
GUT}$.} \vskip 0.3cm

Furthermore, when we introduce a Higgs triplet ${\bf 3}_H =(H_D,
H_S)$ including a doublet with $Y=-1/2$ and a singlet with $Y=1$
in the bulk, only the Higgs doublet remains massless by the
boundary conditions with the same $(Z,T)$ as for the gauge field :
\begin{equation}
{\bf 3}_H (y)=Z {\bf 3}_H (-y)=T {\bf 3}_H (y+\pi R)
\end{equation}
But, for Type I, when the lepton triplet resides at the fixed
point $O$ where $SU(3)_W$ is fully operative, it is impossible to
have realistic charged lepton masses with a Higgs triplet via
${\bf 3}_\ell{\bf 3}_\ell{\bf 3}_H$. Thus, in that case, we
instead should introduce a Higgs sextet $\bar{\bf 6}_H
=(H_D,H_T,H_S)$ including a doublet with $Y=-1/2$, a triplet with
$Y=1$ and a singlet with $Y=-2$ satisfying the boundary conditions
as
\begin{equation}
\bar{\bf 6}_H (y)=Z \bar{\bf 6}_H (-y) Z^{-1}=T \bar{\bf 6}_H
(y+\pi R) T^{-1}.
\end{equation}
The fixed point $O$ also breaks $SU(3)_W$ down to $SU(2)_L\times
U(1)_Y$. We call it $O_{\rm SM}$ in this case.

We distinguish these two cases by
\begin{eqnarray}
Type\ I: O_{\rm GUT}\ {\rm and\ }A\nonumber\\
Type\ II: O_{\rm SM}\ {\rm and\ }A
\end{eqnarray}

Even in Type II models, quarks should be put at $A$ since quarks
can not fit to the hypercharges of the triplet. Leptons form {\bf
3} of $SU(3)_W$ and can be located at $O$. The schematic behavior
of the symmetries at $O$ is shown in Fig. 1.

In any case, the minimal model with supersymmetry dictates to put
the fields as follows,
 \bea
  O (y=0) & : & L, e^c \nn \\
 {\rm B(ulk)} & : & A_M, H_d \ \ \ (with\ supersymmetry) \\
 A (y=\frac{\pi}{2} R) & : & Q,
 u^c, d^c, H_u \nn
 \eea

For nonsupersymmetric case, one Higgs is enough for the generation
of masses in its minimal form and two Higgs doublet (2HD) model
involves second Higgs. However, for supersymmetric case, two
Higgses are necessary but the necessary positions are different.
At $O$, we need $H_d$ to give masses to charged leptons. At $A$,
we need $H_u$ and $H_d$ to give masses to up type and down type
quarks. From this, we can conclude that $H_u$ is necessary at $A$
and $H_d$ is necessary at $O$ and $A$. Therefore, the minimal
setup is to have $H_u$ at $A$ and $H_d$ in the bulk.

In this paper, therefore, we study following models,\\

\noindent
{\bf Model A}: Scherk-Schwarz gauge symmetry breaking without
supersymmetry\\

We consider Type I with $O_{GUT}$ and $A$ for nonsupersymmetric
cases. First, one Higgs model (A-1) is summarized which has been
studied already. Two Higgs doublet Standard Model is realized in
5D as a vectorlike model (A-2v) and a chiral model (A-2c).\\

\noindent
{\bf Model B}: Scherk-Schwarz gauge symmetry breaking
with supersymmetry\\

At the scale $1/R$, the gauge symmetry is broken but N=1
supersymmetry is unbroken\footnote{Supersymmetry is assumed to be
broken at $M_Z$ or 1 TeV at one fixed point and gaugino mediates
supersymmetry breaking to the opposite(or SM) fixed point. If
leptons are at the fixed point where supersymmetry is broken,
flavor changing neutral currects are induced generally. Thus,
phenomenologically, all the SM matter fields are required to be at
$A$ if supersymmetry is broken at $O$.}. Supersymmetry requires
two Higgs doublets. We review vectorlike realization (B-v) and
chiral realization (B-c) of MSSM in 5D. In general, the bulk
supersymmetry with eight supercharges restricts possible
interactions of the theory and the model is highly predictive
except the Fayet-Iliopoulos(FI) term allowed at the fixed points.
The FI term is
\begin{eqnarray}
{\cal L}_{FI} = \xi_1 D \delta (y) + \xi_2 D \delta (y-\frac{\pi
R}{2}).
\end{eqnarray}
If N=1 is unbroken, the integrated FI term ($(\xi_1+\xi_2) D $)
does not get any radiative correction for $U(1)_Y$ since the FI
term is generated only for anomalous $U(1)$'s. Therefore,
$\xi_1+\xi_2=0$ is a stable setup from the naturalness argument.
For Type I, there is no $U(1)$ at $O_{GUT}$ and $\xi_1 =0$.
Therefore, combining two facts (unbroken nonabelian gauge group at
one fixed point $O_{\rm GUT}$ and 4D N=1 supersymmetry) tells us
that the absence of FI term at $A$ ( $\xi_2=0$) is stable against
radiative corrections. This is one of the merits which allow the
analysis simple. For Type II, quadratically divergent FI terms are
generated at both fixed points with opposite sign and this makes
the configuration complicated and
unpredictable\footnote{Localization of bulk fields by the VEV of
(parity-odd) adjoint chiral fields \cite{nilles} and correction to
the gauge coupling running through Chern-Simons term \cite{CKK}
are two major effects of FI terms.} Though the physics related to
the Fayet-Iliopoulos term is interesting by itself, it goes beyond
the scope of this paper and we consider only Type I in the
paper.\\

\noindent {\bf Model C}: Scherk-Schwarz breaking of both gauge
symmetry and supersymmetry\\

Here, the compactification breaks gauge symmetry and supersymmetry
at the same time. If supersymmetry is also broken by
Scherk-Schwarz mechanism, the Fayet-Iliopoulos term is generated
at $A$ even for Type I since the supersymmetric condition
$\xi_1+\xi_2=0$ for one loop correction disappears and $\xi_1=0$
cannot determine the stable value for $\xi_2$. This makes the
analysis much more complicated and we do not consider
Model C in this paper.\\

Running of gauge couplings is given by \cite{ckk} \bea
\frac{1}{g^2 (M_Z)} & = & \frac{1}{g_3^2} - \frac{b_g}{8\pi^2}
\log \frac{M_{c'}}{M_Z} - \frac{\tilde{b}_g}{8\pi^2} \log
\frac{M_s}{M_{c'}} \nn \\ \frac{1}{{\g}^2 (M_Z)} & = &
\frac{1}{g_3^2} - \frac{b_{\g}}{8\pi^2} \log \frac{M_{c'}}{M_Z} -
\frac{\tilde{b}_{\g}}{8\pi^2} \log \frac{M_s}{M_{c'}} \nn \eea
Here we neglect nonuniversal brane gauge kinetic terms by the
strong coupling assumption and the (moderate) largeness of the
extra dimension. Strong coupling behavior of 5-D gauge coupling
fixes the ratio $M_s/M_{c'}$ to be ${\cal O} (100)$ ($100$ or
$16\pi^3 \sim 500$).

It is convenient for later uses to define the relevant combination
of the beta functions in determining the unification scale.
\begin{eqnarray}
B =  b_g - \frac{b_{\g}}{3}, \tilde{B}  =  \tilde{b}_g -
\frac{\tilde{b}_{\g}}{3}. \nn
\end{eqnarray}
\vskip 0.5cm
\centerline{\bf Model A-1: Standard Model (SM) with
one Higgs doublet}
\vskip 0.3cm

The beta functions below the compactification scale are $ b_g  =
\frac{19}{6}, b_{\g} = -\frac{41}{6}$ and  $B = \frac{49}{9}$.
Above the compactification scale, the beta functions are modified:
$\tilde{b}_g  = -\frac{1}{4}, \tilde{b}_{\g}  = -\frac{27}{4}$ and
$\tilde{B}  = 2$. Useful formula is \cite{HN} \bea \sin^2 \theta_W
& = & \frac{1}{4} - \frac{3}{8\pi} \alpha_{\rm em} \left[
(\tilde{b}_g - \frac{\tilde{b}_{\g}}{3} ) \log \frac{M_s}{M_{c'}}
+ ( b_g - \frac{b_{\g}}{3} ) \log \frac{M_{c'}}{M_Z} \right] \nn
\\ & = & \frac{1}{4} - \frac{3}{8\pi} \alpha_{\rm em} \left[ \tilde{B} \log
\frac{M_s}{M_{c'}} +  B \log \frac{M_{c'}}{M_Z} \right]. \nn \eea
>From this formula, once we fix $\frac{M_s}{M_{c'}} \approx 100$ or
$16 \pi^3$ from naive dimensional analysis, following strong
coupling assumption, we can determine $M_{c'}$ by putting beta
function coefficients for different models. For $M_s/M_{c'} = 100
(16\pi^3)$, the unification scale is given as $70 (190)$ TeV which
is just two to three orders higher compared to the electroweak
scale.

\vskip 0.5cm
\centerline{\bf Model A-2v: Two Higgs doublet SM with
vectorlike embedding}
\vskip 0.3cm

The beta functions below the compactification scale are $ b_g  =
3, b_{\g} = -\frac{21}{3}$ and $B = \frac{16}{3}$. Above $1/R$,
$\tilde{b}_g = (23/6,-1/6,-4)$ for gauge, Higgs and matters,
respectively, and $\tilde{b}_{\g} = (0,-1/6,-20/3)$. Total sum is
$\tilde{b}_g  =  -\frac{1}{3} , \tilde{b}_{\g}  = -\frac{41}{6}$
and $\tilde{B}  = \frac{35}{18}$. Then we can determine the
unification scale $M_s$ and it is $80 (210)$ TeV for $M_s/M_{c'} =
100 (16\pi^3)$.

\vskip 0.5cm
\centerline{\bf Model A-2c: Two Higgs doublet SM with
chiral embedding}
\vskip 0.3cm

This is the minimal model for Higgs configuration. From the
minimality condition, leptons are at $O_{GUT}$ and quarks are at
$A$, then $H_d$ should be in the bulk and $H_u$ should be at $A$
for the two Higgs doublet SM in which both Higgses play roles in
giving fermion masses. Below $1/R$, the beta function is the same
as Model A-2v. Above $1/R$, $\tilde{b}_g = (23/6,-1/4,-4)$ for
gauge, Higgs and matters, respectively, and $\tilde{b}_{\g} =
(0,-1/4,-20/3)$. The sum is $\tilde{b}_g = -\frac{5}{12},
\tilde{b}_{\g} = -\frac{83}{12}$ and $\tilde{B}  = \frac{17}{9}$.
The unification scale is then $80 (220)$ TeV for $M_s/M_{c'} = 100
(16\pi^3)$.

\vskip 0.5cm
\centerline{\bf Model B-v: MSSM with vectorlike
embedding}
\vskip 0.3cm

In this model both $H_d$ and $H_u$ are put in the bulk.
Phenomenological constraint requires $H_d$ from $\bar{\bf 6}_H$
rather than ${\bf 3}_H$ to give correct charged lepton masses.
However, the contribution of $\bar{\bf 6}_H$ and ${\bf 3}_H$ to
the beta function is the same. The reason is following. The
members of the hypermultiplet $\bar {\bf 6}_H=\{H_D,H_T,H_S,\hat
H_S,\hat H_T, \hat H_D \}$ are assigned the $Z_2\times Z_2^\prime$
parity as $\{(++),(+-),(+-),(-+),(-+),(--)\}$. Likewise, for ${\bf
3 }_H=\{H_D,H_S,\hat H_S,\hat H_D\}$, the parity assignments are
$\{(++),(+-),(-+),(--)\}.$ Since only the even modes with $(++)$
and $(--)$ contribute to the logarithmic running, $\bar{\bf 6}_H$
and ${\bf 3}_H$ give the same contribution to the beta function.

The running interval for the supersymmetric models can be divided
into three parts. From $M_Z$ to $M_{\rm SUSY}$, the running is
governed by SM. From $M_{\rm SUSY}$ to $M_{c'}$, the running is
that of the usual MSSM. From $M_{c'}$ to $M_s$, we follow the
analysis given in \cite{kkl}. Though $M_{\rm SUSY}$ can vary from
$M_Z$ to $1$ TeV, we identify $M_{\rm SUSY}$ with $M_Z$ for the
simple analysis. Dependence of the unification scale on the
detailed sparticle spectrum is quite strong and the result should
be understood as a qualitative one with uncertainties from
undertermined sparticle spectrum. The beta function coefficients
are the following. Below $1/R$, $b_g  =  -1, b_{\g} = -11$ and $B
= \frac{8}{3}$. Above the compactification scale,
$\tilde{b}_g=(4,0,-6)$ for gauge, Higgs and matters, respectively,
and $\tilde{b}_{\g}=(0,0,-10)$. The sum is $\tilde{b}_g = -2,
\tilde{b}_{\g}  =  -10$ and $\tilde{B}  = \frac{4}{3}$. The
unification scale is determined to be $ 1.9 (4.2) \times 10^4$ TeV
for $M_s/M_{c'} = 100 (16\pi^3)$. This unification scale becomes
slightly lower if we change $M_{\rm SUSY}$ to 1 TeV from $M_Z$ but
is still highler compared to the nonsupersymmetric models.

\vskip 0.5cm
\centerline{\bf Model B-c (minimal model): MSSM with
chiral embedding}
\vskip 0.3cm

In this model, $H_d$ is put in the bulk and $H_u$ is put at the
fixed point $A$. From the minimality condition, leptons are at
$O_{GUT}$ and quarks are at $A$. $H_d$ should couple at both fixed
points in order to give mass to charged leptons and down type
quarks and $H_u$ is necessary only at $A$. Therefore, Model B-c
realizes the minimal configuration required by the
phenomenological constraint. The running below $1/R$ is the same
as Model B-v. Above the compactification scale,
$\tilde{b}_g=(4,0,-13/2)$ for gauge, Higgs and matters,
respectively, and $\tilde{b}_{\g}=(0,0,-21/2)$. $\tilde{b}_g  =
-\frac{5}{2}, \tilde{b}_{\g}  = -\frac{21}{2}$ and $\tilde{B}  =
1$. Now the unification scale is a little bit higher than the
Model B-v, $M_s = 3.4 (9.2) \times 10^4$ TeV for $M_s/M_{c'} = 100
(16\pi^3)$.

In the previous analysis, the threshold corrrections are neglected
and the possible effects from nonuniversal brane kinetic terms at
$A$ are assumed to vanish. The suppression of the brane kinetic
term is justified by the strong coupling assumption once the size
of the extra dimension is larger than the Planck length (or string
length $l_s = 1/M_s$).

\section{Proton decay}

Longevity of proton at this moment gives strong constraint on the
models with low fundamental scale. Extremely many operators
consistent with the symmetries of the theory should be forbidden
in order to avoid rapid proton decay for low fundamental scale
models. Though it is possible to forbid all these unwanted B
violating operators by imposing discrete gauge symmetry, the
geometric explanation for the suppression of B violating operators
\cite{as} is much simpler and nicer. Once the quarks and leptons
are at different positions along the extra dimension, the B
violating operators are forbidden by the locality of higher
dimensional field theory. Nonlocal B violating terms are generated
by nonperturbative or quantum gravitational effects at low energy
but in the models considered above it is suppressed by the
separated length as $e^{-\frac{M_s}{M_{c'}}} \sim e^{-100} \sim
10^{-44} $ since quarks and leptons are maximally separated in the
extra dimension. Exponential suppression of unwanted B violating
operators is easily obtained once we separate quarks and leptons a
few tens times the fundamental length scale. The minimality
condition of $SU(3)_W$ puts quarks at $A$ and leptons at $O_{GUT}$
and the proton decay is naturally avoided by the minimality
condition.\footnote{More precisely, leptons can be placed in the
bulk and can have contact interations at $A$ with quarks.
Therefore, we should assume that matter chiral fields live only at
fixed points. In this case, different locations of leptons and
quarks are derived from the minimality condition. The presence of
anomalous $U(1)$ under which leptons are charged allows leptons to
be localized at the fixed point via Fayet-Iliopoulos term.
However, in this case the separation of leptons and quarks are not
guaranteed.}

\section{Neutrino mass}

In order to give mass to neutrinos, $H_u$ also should act at $O$.
Thus now we have to consider two bulk Higgses. However, \bea {\cal
L} & = & \frac{\lambda}{M_*} 3_L 3_L \bar{3}_{H_u} \bar{3}_{H_u}
\delta(y) \nn \\ & = & \frac{\lambda}{M_*} L L H_u H_u \delta(y) +
\cdots \eea gives too large masses to neutrinos unless $M_*$ is
extremely higher than the electroweak scale ($\sim \langle H_u
\rangle$). By giving large kink mass we can make $H_u$ almost
localize at $A$ and this explains the smallness of neutrino masses
through the exponential suppresion of $H_u$ wave function. The
detailed realization for the $H_u$ localization needs extra
modification of the models. $U(1)_Y$ cannot play an asymmetric
localization since the hypercharge of $H_u$ and $H_d$ is exactly
opposite and if $H_u$ is localized at one fixed point and then
$H_d$ is localized at the other fixed point. Therefore, we should
introduce anomalous $U(1)_A$ under which $H_u$ and $H_d$ are
asymmetrically charged. (For instance, only $H_u$ is charged under
$U(1)_A$.)

\section{Fixed point anomaly cancellation}

Since we deal with chiral models, it is necessary to justify how
the cancellation of the local gauge anomaly at the orbifold fixed
point is realized in detail.

Anomaly is an IR(infrared) property of the theory and is
determined from the massless sector of the model in 4D. However,
there can exist a local gauge anomaly spread along the extra
dimension if we spread quarks and leptons at different fixed
points even when the integrated anomaly cancels in the effective
4D theory. Therefore, it is necessary to check whether the fixed
point gauge anomaly can be cancelled by writing down local counter
terms.

Firstly, let us briefly review the result on the abelian gauge
anomaly. For a 5D orbifold, the local gauge anomaly appears only
at the fixed point and is equally distributed for an abelian gauge
group. For $U(1)$ gauge group with one unit-charged fermion in the
bulk on an $S^1/Z_2$ orbifold, the local gauge anomaly
is\cite{acg},
\begin{eqnarray}
\partial_M J^M (x,y) & = & \frac{1}{2} \left[ \delta(y) +
\delta(y-\pi R) \right] {\cal Q},
\end{eqnarray}
where $J^M$ is the five dimensional current and
\begin{eqnarray}
{\cal Q} & = & \frac{1}{32\pi^2} F_{\mu\nu} \tilde{F}^{\mu\nu}
\end{eqnarray}
is the four dimensional chiral anomaly. Starting from a theory
with bulk fermions, we can calculate the contributions of all the
Kaluza-Klein modes to the anomaly and the answer is independent of
the wave functions. The chiral fermion contribution localized at
the fixed point is
\begin{eqnarray}
\partial_{\mu} J^{\mu} (x,y) & = & \delta(y) {\cal Q},
\end{eqnarray}
and the Chern-Simons term contribution is
\begin{eqnarray}
\partial_5
J^5 (x,y) & = & \frac{1}{2} \left[ -\delta(y) + \delta(y-\pi R)
\right] {\cal Q}
\end{eqnarray}
from
\begin{eqnarray}
{\cal L}_{\rm CS} = -\frac{1}{128\pi^2} \epsilon(y)
\epsilon_{MNPQR} A^M F^{NP} F^{QR}.
\end{eqnarray}
where $\epsilon(y)=\pm 1$ for $y>0$ and $y<0$, respectively.

The above calculation can be extended to $S^1/ (Z_2 \times
Z_2^{'})$ \cite{zwirner}. Since the five dimensional Dirac spinor
has left-- and right-- handed spinors, $\psi_L,\psi_R$, from the
4D point of view, there are two possibilities for the parity
assignment of the fermions,\vskip 0.2cm

Case (1) $\psi_L$ : $(+,+)$ , $\psi_R$ : $(-,-)$

Case (2) $\psi_L$ : $(+,-)$ , $\psi_R$ : $(-,+)$ \vskip 0.2cm

For Case (1) with the parity $(+,+)$ and $(-,-)$, the anomaly
calculation is the same as the previous result and the local gauge
anomaly is
\begin{eqnarray}
\partial_M J^M (x,y) & = & \frac{1}{2}
\left[ \delta(y) +
\delta(y-\frac{\pi R}{2}) \right] {\cal Q}.
\end{eqnarray}
For the second case with $(+,-)$ and $(-,+)$, the flip of the
parity under $Z_2^{'}$ can be represented by the twisting of the
corresponding fields under the translation $y \rightarrow y + \pi
R$ with $e^{i \frac{y}{R}}$ for the corresponding wave functions.
Therefore, the answer is \cite{zwirner}
\begin{eqnarray}
\partial_M J^M (x,y) & = &  \frac{1}{2} e^{2i \frac{y}{R}}
\left[ \delta(y) +
\delta(y-\frac{\pi R}{2}) \right] {\cal Q}, \nn \\
& = & \frac{1}{2} \left[ \delta(y) - \delta(y-\frac{\pi R}{2})
\right]{\cal Q} .
\end{eqnarray}

It is easy to see that this local gauge anomaly can be cancelled
exactly by the bulk Chern-Simons term,
\begin{eqnarray}
{\cal L}_{\rm CS} = -\frac{1}{128\pi^2} \epsilon(y)
\epsilon_{MNPQR} A^M F^{NP} F^{QR}.
\end{eqnarray}
Now, it is quite straightforward to generalize the above review on
the abelian gauge anomaly cancellation to the nonabelian case.

\subsection{Nonabelian anomaly from bulk matters : unbroken case}

When the gauge symmetry is not broken, the previous formula is
valid. For instance, the anomaly for the fermion $N$ of $SU(N)$
 with the parity $(+,(-1)^{h_i})$ and $(-,-(-1)^{h_i})$) is \bea
D_M J^{Ma} & = & \frac{1}{2} \left[ \delta(y) + (-1)^{h_i}
\delta(y-\frac{\pi R}{2}) \right] {\cal Q}^a, \eea where \bea
{\cal Q}^a & = & \frac{1}{32\pi^2} D^{abc} F^b_{\mu\nu}
\tilde{F}^{c\mu\nu}, \eea and \bea D^{abc} & = & \frac{1}{2} {\rm
tr} \left( \{T^a,T^b\}T^c \right). \eea If the fermion contains
massless zero mode ($h_i=0$), the anomaly appears at both fixed
points with equal sign, and if it does not have massless zero mode
($h_i=1$) , the anomaly appears with opposite sign. This is a
trivial generalization of the abelian result.

\subsection{Nonabelian anomaly cancellation with
Scherk-Schwarz breaking of gauge symmetry}

However, the most interesting models involve the breaking of gauge
symmetry with the parity. Then the symmetric fixed point $O_{GUT}$
and the other fixed point $A$ should be distinguised. For the bulk
gauge group $G$, the $Z_2^{'}$ parity which does not commute with
$G$ break the gauge group to $H = H_1 \times H_2 \times \cdots
\times H_n$. The anomaly at $O_{GUT}$ is expressed $G$ invariantly
and is the same as the above. At $A$, the expression needs more
information.

\subsubsection{Bulk fermion contribution}

Bulk fermions belonging to the fundamental representation of $G$
is divided into fundamentals under $H_i$ and have different
parities under $Z_2^{'}$. Let the parity be $(+,(-1)^{h_i})$ and
$(-,-(-1)^{h_i})$ for $\psi_L$ and $\psi_R$. Then we can calculate
the anomaly induced at $A$ from the knowledge of abelian gauge
anomaly just like the previous section except the fact that now
$\cal Q$ is not the anomaly of entire bulk gauge group but is that
of unbroken subgroup. Thus the general expression of the anomaly
from the bulk matter is
\begin{eqnarray}
D^M J_M^{a} & = & \frac{1}{2} \delta(y) {\cal Q}_O^a + \frac{1}{2}
\delta(y-\frac{\pi R}{2}) \sum_i (-1)^{h_i} {\cal Q}_{Ai}^{a_i}
\delta_{a a_i},
\end{eqnarray}
where
\begin{eqnarray}
{\cal Q}_O^a & = & \frac{1}{32\pi^2} D^{abc} F^b_{\mu\nu}
\tilde{F}^{c\mu\nu},
\end{eqnarray}
and
\begin{eqnarray}
{\cal Q}_{Ai}^{a_i} & = & \frac{1}{32\pi^2} D^{a_i b_i c_i}
F^{b_i}_{\mu\nu} \tilde{F}^{c_i\mu\nu}.
\end{eqnarray}
Here, $a$ is the gauge index for unbroken group $G$ and $a_i$ is
the gauge index of its subgroup $H_i$.

\subsubsection{Chern-Simons contribution}

The bulk Chern-Simons term is
\begin{eqnarray} {\cal L}_{CS} & = & -\frac{1}{128\pi^2} \epsilon(y)
{\rm tr} \left( A F^2 - \frac{1}{2} A^3 F + \frac{1}{10} A^5
\right).
\end{eqnarray}
The gauge transformation leaves a nonvanishing term at the
boundary which can cancel the anomaly from the chiral fermions.
The general expression of the anomaly from the Chern-Simons term
is
\begin{eqnarray}
D^M J_M^a & = & -\frac{1}{2} \delta(y) {\cal Q}_O^a + \frac{1}{2}
\delta(y-\frac{\pi R}{2}) \sum_i {\cal Q}_{Ai}^{a_i} \delta_{a
a_i},
\end{eqnarray}
where
\begin{eqnarray}
{\cal Q}_O & = & \frac{1}{32\pi^2} D^{abc} F^b_{\mu\nu}
\tilde{F}^{c\mu\nu}, \eea and \bea {\cal Q}_{Ai} & = &
\frac{1}{32\pi^2} D^{a_i b_i c_i} F^{b_i}_{\mu\nu}
\tilde{F}^{c_i\mu\nu}.
\end{eqnarray}

\subsubsection{Bulk fermion, Chern-Simons term, and brane fermion
contributions}

If we add two contributions, the anomaly at $O_{GUT}$ cancel with
each other perfectly. The anomaly at $A$ is
\begin{eqnarray}
D^M J_M^a & = & \delta(y-\frac{\pi R}{2}) \sum_{i} {\cal
Q}_{Ai}^{a_i} \delta_{a a_i} \delta_{h_i 0} .
\end{eqnarray}
The contributions of bulk fermions with $h_i=1$ and Chern-Simons
term cancel while the contributions of those with $h_i=0$ and CS
term add up. This is the anomaly that can be cancelled exactly by
putting incomplete multiplet associated to the zero mode which is
(anti-)fundamental under $H_i$ (or carries opposite charges
compared to the zero mode from the bulk fermion). This clearly
shows the cancellation mechanism of cubic gauge anomaly like
YM-YM-YM, $U(1)$-YM-YM and $U(1)^3$.

\subsection{$U(1)$ gauge boson--graviton--graviton anomaly
cancellation}

In the previous sections, we studied the cubic anomaly
cancellation of abelian and nonabelian gauge groups. In this
subsection, the mixed anomaly for 
$U(1)$ gauge boson--graviton--graviton is
considered. For an abelian gauge group, $U(1)$--graviton--graviton
anomaly cancellation is done in parallel to $U(1)^3$ anomaly
cancellation. At each fixed point the anomaly is
\begin{eqnarray}
\partial_M J^M (x,y) & = & \frac{1}{2} \left[ \delta(y) +
\delta(y-\frac{\pi R}{2}) \right] {\cal Q_G},
\end{eqnarray}
where $J^M$ is the five dimensional current and
\begin{eqnarray}
{\cal Q_G} & = & \frac{1}{192\pi^2} R_{\mu\nu} \tilde{R}^{\mu\nu}
\end{eqnarray}
is the gravitational anomaly. The chiral fermion localized at the
fixed point gives
\begin{eqnarray}
\partial_{\mu} J^{\mu} (x,y) & = &
\delta(y) {\cal Q_G}'
\end{eqnarray}
and the Chern-Simons term gives
\begin{eqnarray}
\partial_5 J^5 (x,y) & = & \frac{1}{2} \left[ -\delta(y) +
\delta(y-\frac{\pi R}{2}) \right] {\cal Q_G},
\end{eqnarray}
where
\begin{eqnarray}
{\cal L}_{\rm CS} = -\frac{1}{768\pi^2} \epsilon(y)
\epsilon_{MNPQR} A^M R^{NP} R^{QR}.
\end{eqnarray}
We obtain a similar conclusion for $S^1/Z_2 \times Z_2^{'}$.

The apparent problem appears when a $U(1)$ gauge group survives
after the breaking of the bulk nonabelian gauge group. In this
case, the gravitational mixed anomaly of $U(1)$ is induced only at
the fixed point $A$ and should be cancelled by localized fields at
$A$ since the bulk has nonabelian gauge symmetry and does not
allow a gravitational Chern-Simons term like $A \wedge R \wedge
R$. To see how this works, let us consider the breaking of
$SU(M+N)$ to $SU(M) \times SU(N) \times U(1)$ by Scherk-Schwarz
mechanism at $A$. If the fundamental $M+N$ of $SU(M+N)$ is in the
bulk has a parity assignment $(+,-)$ and $(-,+)$ for $M_{aN}$ and
$(+,+)$ and $(-,-)$ for $N_{-aM}$, there appears a $U(1)$
gravitational mixed anomaly at $A$
\begin{eqnarray}
\partial_M J^M (x,y) & = & \frac{1}{2} \left[ (-1)M\times{aN} +
N\times(-aM) \right] \delta(y-\frac{\pi R}{2}) {\cal Q_G},
\end{eqnarray}
where the subscript $aN$ and $-aM$ denote $U(1)$ charge and we
haven't fixed the normalization of $U(1)$ charge $a$. Since the
$U(1)$ is a subgroup of simple group, the sum of $U(1)$ charges
should vanish. Therefore the anomaly becomes
\begin{eqnarray}
\partial_M J^M (x,y) & = &
N\times(-aM)  \delta(y-\frac{\pi R}{2}) {\cal Q_G},
\end{eqnarray}
and is cancelled by the $U(1)$--graviton--graviton anomaly induced
from $\bar{N}_{aM}$ localized at $A$. This result is very
interesting since we cannot write down YM-gravity-gravity mixed
Chern-Simons term $A \wedge R \wedge R$ due to ${\rm Trace}(A)
=0$. The $U(1)$ gravitational mixed anomaly should be cancelled
without the aid of Chern-Simons term and this is the case indeed.

This is in accord with the absence of the Fayet-Iliopoulos term in
the setup. $N=1$ supersymmetry relates $U(1)$ gravitational mixed
anomaly from the fermions to the FI term from the bosons, and the
absence of the anomaly guarantees the absence of quadratically
divergent FI term at the fixed point. Therefore, the vanishing of
FI term is natural and is protected from radiative corrections.

\subsection{$SU(5)$}

Even though we are studying the $SU(3)_W$ model here, it is
appropriate to see the cancellation of the anomalies for the
chiral model\cite{kkl} since we presented above the general
mechanism for the fixed point anomaly cancellation. In the bulk
there exists one ${\bf 5}$ with parity $(-,-,-,+,+)$ and one
doublet ${\bf 2}$ with opposite hypercharge is locate at $A$.
$H_d$ is the zero mode coming from ${\bf 5}$ and is ${\bf
2}_{-\frac{1}{2}}$. $H_u$ located at $A$ is ${\bf
2}_{\frac{1}{2}}$. The cubic anomaly of $SU(5)$ from ${\bf 5}$
appears at $O$ and should be cancelled by the Chern-Simons term.
Then at $A$, the cubic anomaly of the subgroup $SU(3)_C^3$ and
$U(1)_Y^3$ (and also the mixed anomaly $U(1)_Y-SU(3)_C-SU(3)_C$
from the triplet of ${\bf 5}$ and the Chern-Simons term cancel
with each other. The cubic anomaly $SU(2)_L^3$ and $U(1)_Y^3$ (and
the mixed anomaly $U(1)_Y-SU(2)_L-SU(2)_L$) from the doublet of
$5$ is added with the one from the Chern-Simons term and is
cancelled by the anomaly from the doublet living at $A$. The
gravitational mixed anomaly of $U(1)$ from ${\bf 5}$ can appear at
$A$ as
\begin{eqnarray}
\partial_M J^M (x,y) & = &
1/2 \left[ (-1) \times 3 \times \frac{1}{3} + 2 \times
(-\frac{1}{2}) \right] \delta(y-\frac{\pi R}{2}) {\cal Q_G} \nn \\
& = & (-1) \delta(y-\pi R) {\cal Q_G} ,
\end{eqnarray}
where the first term is from ${\bf 3}_{\frac{1}{3}}$ and the
second is from ${\bf 2}_{-\frac{1}{2}}$. This is exactly cancelled
by $U(1)_Y$-gravity-gravity anomaly from ${\bf 2}_{\frac{1}{2}}$
living at $A$.

\subsection{$SU(3)_W$}

The anomaly of Higgs sector cancels independently and quark-lepton
sector cancels with appropriate Chern-Simons terms. In the
previous discussions it was proven that once the 4D gauge anomaly
cancel the 5D fixed point gauge anomaly can be cancelled with
having appropriate Chern-Simions term for the Scherk-Schwarz
breaking setup of gauge symmetry.

Considering $T_8=-Y/\sqrt{3}$, we find that the fixed point
anomalies from leptons ${\bf 3}_l$ appear only at $O_{GUT}$ and
are cancelled by the 5D Chern-Simons term,
\begin{equation}
{\cal L}_{CS}=-\frac{c_l}{128\pi^2}\epsilon(y) {\rm tr}\ (A\wedge
F\wedge F-\cdots),
\end{equation}
by choosing $c_l=2$ ($3c_l=6$ for three generations) since the
anomaly from the fixed point localized matter is twice of that of
bulk matter. This Chern-Simons term at the same time cancel the
anomalies at $A$ from quarks. Moreover, for the supersymmetric
$SU(3)_W$ case, the anomaly from the bulk Higgsino ${\bf
\bar{6}}_H$ is the same as that of ${\bf 3}$ and appear at
$O_{GUT}$ and $A$. The anomaly at $O_{GUT}$ from ${\bf \bar{6}}_H$
is cancelled by the Chern-Simons term with $c_H=1$, and the
anomalies at $A$ from ${\bf \bar{6}}_H$ and the Chern-Simons term
are cancelled by the contribution from the Higgs doublet ${\bf
2}_{H_u}$ at $A$. Thus, if $ c =3c_l+c_H $ is chosen to be $7$,
all the fixed point anomaly from quarks, leptons and Higgsinos
disappear. For the model in which both Higgses are in the bulk,
$c=3c_l=6$ is enough to cancel all the fixed point anomalies.

Gravitational mixed anomaly of $U(1)_Y$ appears only at $A$.
Quarks do not give the anomaly since $\sum Y_{q_i}=0$. ${\bf
\bar{6}}_H$ gives the anomaly at $A$ as
\begin{eqnarray}
\partial_M J^M (x,y) & = &
1/2 \left[ (-1) \times 1 \times (-2) + 2 \times (-\frac{1}{2}) +
(-1) \times 3 \times 1 \right] \delta(y-\frac{\pi R}{2}) {\cal
Q_G} \nn
\\ & = & (-1) \delta(y-\pi R) {\cal Q_G} ,
\end{eqnarray}
where the first term is from ${\bf 1}_{-2}$ and the second is from
${\bf 2}_{-\frac{1}{2}}$ and the third term is from ${\bf 3}_1$.
Notice $(-1)$ factors for the singlet and the triplet due to the
parity assignment. This is just twice of the bulk doublet
contribution and is exactly cancelled by $U(1)_Y$-gravity-gravity
anomaly from ${\bf 2}_{\frac{1}{2}}$ living at $A$.

\section{Conclusion}

In this paper we examined the possibility of unification of
electroweak gauge group into $SU(3)_W$ with minimality condition.
For the simplest setup, the tree level prediction is very close to
the observed value and the unification is achieved near TeV. As a
specific model, $\lq$natural' location of matters are assumed such
that leptons are at $O$ and quarks are at $A$. The minimal
configuration requires $H_d$ at both fixed points but $H_u$ only
at one fixed point $A$ if we neglect neutrino masses. The setup
can avoid proton decay and can explain the bottom-top mass ratio
with order one $\tan \beta$ as a byproduct of the minimality
condition. Neutrino masses are easily incorporated once we put
$H_u$ also in the bulk. Once kink mass of $H_u$ is given, then
$H_u$ is almost localized at $A$ and neutrinos acquire tiny mass
even for a very low fundamental scale. The vectorlike and chiral
models we considered in this paper are shown to be made free of
anomalies even at the fixed points. From this experience, we
conjecture that the orbifold compactification can be made sensible
even at the fixed points by including an appropriate Chern-Simons
term(axion coupling term with Dirac index density) in the bulk for
odd(even) dimensions if the effective 4D gauge anomalies from the
bulk and fixed points fermions cancel.

\acknowledgments We would like to thank Kiwoon Choi for
helpful discussions. JEK and HML are supported in part by the BK21 program
of Ministry of Education, the KOSEF Sundo Grant, and by the Office
of Research Affairs of Seoul National University.

\end{document}